\documentclass[preprint,preprintnumbers,showpacs,amsmath,amssymb,aps,prb]{revtex4-1}
\usepackage{graphicx}
\usepackage{dcolumn}
\usepackage{bm}

\begin{document}

\title{Metamaterial-Enhanced Coupling between Magnetic Dipoles for Efficient Wireless Power Transfer}

\author{Yaroslav Urzhumov and David R. Smith}
\affiliation{Center for Metamaterials and Integrated Plasmonics,\\
Pratt School of Engineering,\\
Duke University,\\
Durham, N. Carolina, 27708}

\newcommand{\ba}{\begin{eqnarray}}
\newcommand{\ea}{\end{eqnarray}}
\newcommand{\be}{\begin{equation}}
\newcommand{\ee}{\end{equation}}
\newcommand{\para}{\parallel}

\def \d{\partial}
\def \Re{{\rm Re}}
\def \Im{{\rm Im}}
\def \diag{{\rm diag}}
\def \const{{\rm const}}
\def \eff{{\rm eff}}
\def \sign{{\rm sign}}

\begin{abstract}
Non-radiative coupling between conductive coils is a candidate mechanism for wireless energy transfer applications.
In this paper, we propose a power relay system based on a near-field metamaterial superlens,
and present a thorough theoretical analysis of this system. We use time-harmonic circuit formalism to describe all interactions between two coils attached to external circuits and a slab of anisotropic medium with homogeneous permittivity and permeability. The fields of the coils are found in the point-dipole approximation using Sommerfeld integrals, which are reduced to standard special functions in the long-wavelength limit.
We show that, even with a realistic magnetic loss tangent of order 0.1, the power transfer efficiency with the slab can be an order of magnitude greater than free-space efficiency when the load resistance exceeds a certain threshold value. We also find that the volume occupied by the metamaterial between the coils can be greatly compressed by employing magnetic permeability with a large anisotropy ratio.

\end{abstract}


\maketitle

\section{\label{sec:intro}Introduction}

The explosive growth in the use of cordless hand-held electronic devices, electrical vehicles and appliances in the recent years is stimulating interest in wireless power sources~\cite{fernandez_borras01,hamam_soljacic10,chiao11}. Wireless communications between stationary and mobile devices make it possible to remotely control devices, giving them a certain degree of autonomy. However, since all electrical devices depend on an energy supply, it must be either carried with the device, or harvested from the environment as needed.

Energy harvesting can include mechanical, thermal, chemical, gravitational or electromagnetic energy, although only the latter is suitable for most applications. Radiative transfer of electromagnetic energy is limited by absorption and scattering in the atmosphere and requires a direct line of sight between the source and the device~\cite{brown84}. At high intensities, it also presents a challenging electromagnetic interference problem. While in the radiation flux the magnitudes of electric and magnetic fields are always comparable, in the near field of magnetic dipole antennas the $E/H$ ratio can be strongly suppressed, thus reducing the interaction with biological and other environmental objects whose properties are almost purely dielectric. This bio-friendly behavior of inductively coupled resonators is widely used in RFID transponders~\cite{cardullo_parks73}. Thus, non-radiative inductive coupling between high-frequency circuits within their near-field zone is an attractive option for wireless power transfer applications.

Although inductive near-field coupling between small coils of radius $r\ll \lambda$ separated by a distance $d$ such that $r\ll d \ll \lambda$ is generally small -- proportional to $(r/\lambda)(r/d)^3$ -- it can be resonantly enhanced if the coils are either self-resonant or connected to external resonant circuits. Wireless power transfer over a distance of 8 times the radius of the coils with efficiency $40\%$ has been experimentally demonstrated by Soljacic et al.~\cite{kurs_soljacic07,hamam_soljacic09}. However, the power transfer efficiency of their device drops steeply as a function of the distance between the coils, as well as a function of the resistive load attached to either of the coils. To address these issues, we propose a relay system based on the concept of the near-field superlens~\cite{veselago68,pendry_lens00}, which can greatly enhance both the range of distances and the load resistance at which the power transfer efficiency is high enough to be practically useful.

Shortly after the discovery of the negative-index superlens~\cite{veselago68,pendry_lens00,smith_pendry03,pokrovsky_efros03}, it was realized that its superresolution is related to the enhancement of the transfer function of the near fields, in addition to its aberration-free refractive properties~\cite{smith_pendry03,pokrovsky_efros03,dong_chan10}. Reduced near-field-only superlenses based on negative permittivity media were theoretically proposed~\cite{shamonina_el01,ramakrishna02,shvets_spie03,merlin_apl04,blaikie_apl04,shvets_urzh_mrs04,lu_05} and implemented shortly afterwards~\cite{zhang_sci05,taubner_shvets_science06}. An important aspect of negative-$\epsilon$ lenses is that they enhance transmission of only TM-polarized waves (with magnetic field normal to the plane of propagation). In two-dimensional geometries, TM waves can be generated by either in-plane polarized electric dipoles or out-of-plane magnetic line currents. The near-perfect transfer of magnetic, as well as electric, fields has been demonstrated in two-dimensional systems with purely dielectric superlenses~\cite{shamonina_el01,ramakrishna02,shvets_spie03,merlin_apl04,blaikie_apl04,shvets_urzh_mrs04,lu_05,zhang_sci05,taubner_shvets_science06}. Similar results are expected with purely magnetic superlenses~\cite{freire_marques06,freire_lapine10,choi_seo10} due to electromagnetic duality.

In contrast with their two-dimensional counterparts, three-dimensional {\it magnetic} point dipoles are difficult to couple using a {\it dielectric} superlens, because the TM wave component of their magnetic field is suppressed in the near-field relative to the TE wave component by a small factor of order $(d/\lambda)^2$ (see Refs.~\cite{chew94,choi_seo10} and equation~\ref{eq:Hx_axis} below). Fortunately, for applications relying on radio or microwave frequencies, artificial media with negative magnetic permeability have been designed and demonstrated~\cite{pendry_stewart99,smith_schultz00}. Negative-$\mu$ metamaterial lenses have been proposed for improved magnetic resonance imaging~\cite{freire_marques06,freire_lapine10} and wireless energy transfer~\cite{choi_seo10} applications. In this article, we focus on finding the optimum coupling regime that maximizes the amount of energy transferred to the useful load. In performing the analysis, we allow the  metamaterial to be anisotropic. Earlier studies suggested that strongly anisotropic media can serve as enhanced imaging systems and may dramatically enhance the transfer function~\cite{smith_rye_apl04,smith_schurig_josab04,gallina_engheta10,dong_chan10}. Eliminating the need for strong magnetic response in one or more directions is potentially interesting as it could reduce the complexity and cost of such metamaterials. However, our findings indicate that the optimum coupling is obtained in the regime where all three components of $\mu$ tensor are simultaneously negative. On the other hand, simultaneous control over all three principal values of $\mu$ enables reduction of the superlens thickness~\cite{gallina_engheta10,dong_chan10}.

This paper is organized as follows. First, we introduce a circuit model of the inductive coupling between two coils and one metamaterial slab, and define the power transfer efficiency equivalent to the one introduced in Ref.~\cite{kurs_soljacic07}.
Then, we calculate the mutual and self-inductances of the coils in the presence of the slab, and show that a lossy metamaterial introduces resistive contributions to the otherwise purely reactive response of perfectly conducting coils. The coils are described as point dipoles where applicable, and the metamaterial is replaced with an equivalent homogeneous, possibly anisotropic, medium. Then, we analyze the power transfer efficiency and find the optimum coupling regime. Finally, several important aspects of this relay system are discussed: the effect of the dipole orientation relative to the slab, contribution of the TM waves and the possibility of enhancing magnetic fields with a dielectric-only metamaterial, and the role of anisotropy.


\section{\label{sec:circuit} The circuit model of magnetic dipole coupling}

\begin{figure}
\centering
\begin{tabular}{cc}
\includegraphics[width=0.4\columnwidth]{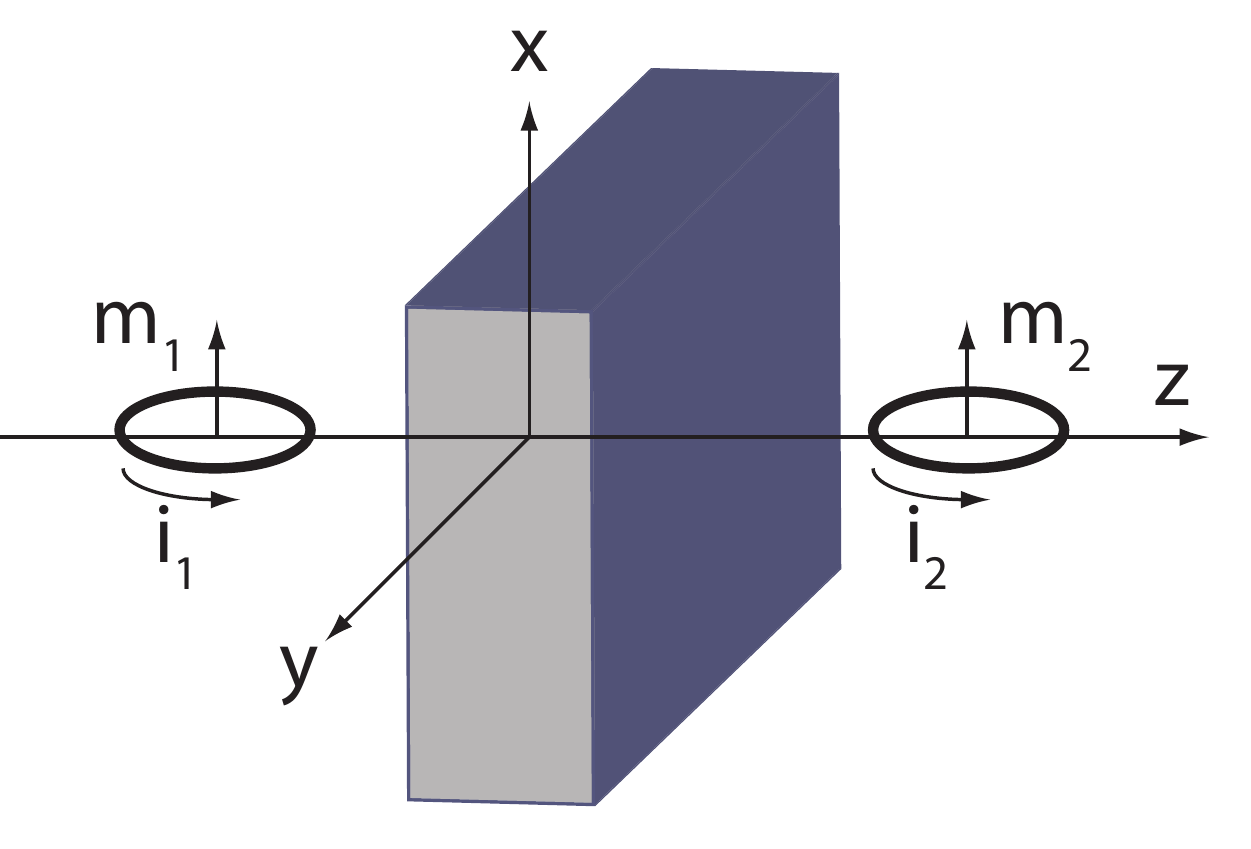}&
\hskip0.1\columnwidth
\includegraphics[width=0.4\columnwidth]{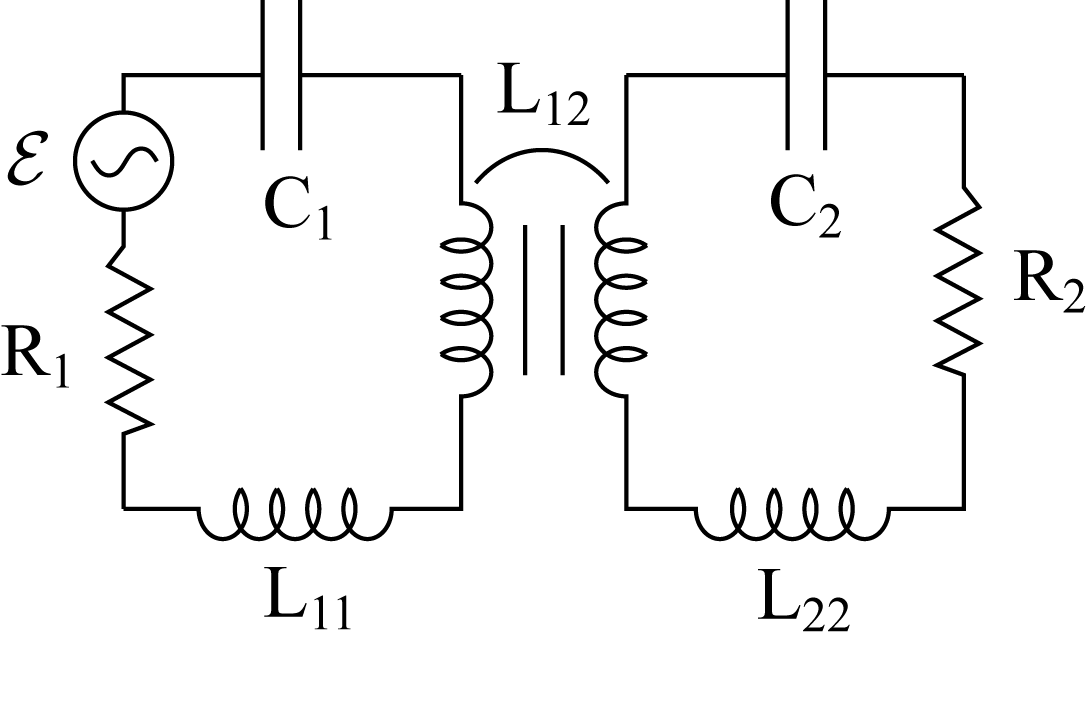}
\\
(a)&(b)\\
\end{tabular}
\caption{
(Color online)
Left: The schematic of the device consisting of two single-turn coils coupled through a slab of homogeneous, possibly anisotropic medium, of thickness $D$. The $z$ axis is normal to the slab, with the origin (0) on the left face of the slab.
The coils are positioned on the $z$ axis, with their magnetic dipole moments oriented in the $x$-direction.
The distance from the center of coil 1 to the left face of the slab is $d_1$, and from coil 2 to the right face of the slab it is $d_2$.
Right: Circuit diagram of the device.
}
\label{fig:coils_slab}
\end{figure}

Near-field coupling between magnetic dipoles is an attractive option for radiation-less power transfer. In contrast with electric dipoles, magnetic field in the vicinity of magnetic dipoles dominates strongly over the electric field; thus, magnetic dipole-based energy transfer system is expected to interfere less with the function of other devices and live biological objects, including humans. A simple implementation of a magnetic dipole -- a circular, planar, single-turn loop of wire -- is assumed in this study (see Fig.~\ref{fig:coils_slab}(a)).
The magnetic moment of a wire loop relates to its current $i$ and area $A=\pi R^2$ as $\vec m_0=iA \vec n$, where $\vec n$ is a unit vector normal to the plane of the loop.

Electromagnetic interaction between two {\it small} wire loops can be described with the aid of the coupled-coil theory~\cite{fano_chu_adler60}, which states that the magnetic flux through coil $m$ is expressible through the currents in all participating coils:

\be \Phi_m \approx (\vec B_m\cdot \vec n_m)A_m = \sum_{n=1}^{2} L_{mn} i_n, \label{eq:coupled-coils-flux}
\ee
where $A_m$ is the area of the $m$-th coil, $\vec B_m$ is magnetic flux density at the center of the loop, and the coefficients $L_{mn}$ are called self-inductances for $m=n$, and mutual inductances for $m\ne n$.
The radius of each coil is assumed to be much smaller than both the free-space wavelength and the spatial scale of variation of the magnetic fields generated by all other coils.
The expressions (\ref{eq:coupled-coils-flux}) remain valid even if the coils are located in the vicinity of objects with strong electromagnetic response, for example, a metamaterial slab, as shown in Fig.~\ref{fig:coils_slab}. Inductances $L_{mn}=L_{nm}$ are affected by the coils environment; they are calculated for the two-coil-and-one-slab system below.

Using Faraday's law of induction, the relationships (\ref{eq:coupled-coils-flux}) can be re-written in terms of the induced electromotive force (EMF), i.e., inductively induced voltage in each coil:

\be V_m = \sum_{n=1}^{2} L_{mn} \frac{di_n}{dt}=-j\omega\sum_{n=1}^{2} L_{mn} i_n. \label{eq:coupled-coils-emf}
\ee
We assume time-harmonic dependence of the currents and fields as $i\propto e^{-j\omega t}$. Assume that coils 1 and 2 are connected to external circuits with resistances $R_m$ and capacitances $C_m$, $m=1,2$; additionally, the circuit of coil 1 includes a voltage generator with a fixed external EMF $\mathcal{E}$. From the two voltage balance equations,
\ba V_1 = \mathcal{E} +i_1 R_1 +\frac{i_1}{j\omega C_1} = -j\omega(L_{11}i_1 + L_{12}i_2),\nonumber \\
V_2 = i_2R_2 +\frac{i_2}{j\omega C_2} = -j\omega(L_{22}i_2 +L_{21}i_1),
\label{eq:V12}
\ea
the currents in the two coils are expressed through the external EMF, giving
\ba i_1 = \frac{-\mathcal{E}}{Z_1^{\eff}},\nonumber \\
i_2 = -j\omega\frac{L_{21}}{Z_2}i_1=j\omega\frac{L_{21}\mathcal{E}}{Z_1^{\eff}Z_2},
\label{eq:i12}
\ea
where we have introduced effective coil impedances
\ba Z_1^{\eff} = Z_1 +\omega^2\frac{L_{12}L_{21}}{Z_2},\nonumber \\
Z_m = R_m+j\omega(L_{mm}-\frac{1}{\omega^2C_m}), m=1,2.
\label{eq:Z12}
\ea

The power dissipated in each circuit is given by the sum of terms $P_m=\frac{1}{2}\Re(V_m i_m^*)=\frac{1}{2}\Re(Z_m^\eff)|i_m|^2$, summed over all circuit elements with non-vanishing resistance; explicitly,
\ba P_1 = \frac{1}{2}R_1^\eff\frac{|\mathcal{E}|^2}{|Z_1^\eff|^2},\nonumber \\
P_2 = \frac{1}{2}R_2^\eff\omega^2\frac{|L_{21}|^2\mathcal{E}|^2}{|Z_1^\eff|^2|Z_2|^2},
\label{eq:P12}
\ea
where we have introduced effective resistances
\ba
R_1^\eff=\Re Z_1^\eff = R_1 - \omega \Im L_{11} + \omega^2 \Im \frac{L_{12}L_{21}}{Z_2},\\
R_2^\eff = \Re Z_2 = R_2 - \omega \Im L_{22} + \omega^2 \Im \frac{L_{12}L_{21}}{Z_1}.
\ea

Note that the self-inductances of coils in the presence of a lossy slab may have a non-zero imaginary part, which accounts for the resistive, dielectric and magnetic losses in the slab. In the following, we will be neglecting the terms quadratic in frequency, which originate from the resistive part of mutual impedance $Z_{12}\equiv j\omega L_{12}$. Because of this resistive part, one should distinguish the power dissipated in the useful resistive load $R_2$,
\be
P_2^{0}=\frac{1}{2}R_2|i_2|^2,
\ee
from the total power dissipated in circuit 2 ($P_2$).

The fraction of power delivered to the resistive load $R_2$ relative to the total consumed power is
\be\eta = \frac{P_2^{0}}{P_1+P_2}=\frac{R_2}{R_2^\eff}\frac{\chi}{1+\chi},
\ee
 where
\be \chi \equiv \frac{P_2}{P_1} = \frac{R_2^\eff\omega^2|L_{21}|^2}{R_1^\eff|Z_2|^2}.
\label{eq:chi_ratio}
\ee
In order to make the power transfer efficiency $\eta$ close to $100\%$, one wants $\chi$ to be as large as possible, and the ratio $R_2^\eff/R_2\ge 1$ as small as possible. To maximize $\chi$, it is beneficial to minimize $|Z_2|$ by eliminating $\Im Z_2$; this is achieved by tuning circuit 2 to its resonance by adjusting the capacitance $C_2$, such that $\Re L_{22}-\frac{1}{\omega^2C_2}=0$. In the following, we assume that $Z_2=R_2^\eff$, and thus,
\be \chi = \frac{\omega^2|L_{21}|^2}{R_1^\eff R_2^\eff}.
\label{eq:chi_ratio_res}
\ee

We observe that the power transfer efficiency $\eta$ is not sensitive to the magnitude of $Z_1^\eff$, but only to its resistive part. That is because the ratio $\chi$, as well as efficiency $\eta$, do not reflect the actual power extracted from the voltage generator $\mathcal{E}$ by the load. In order to characterize the effect of quenching of the source in the presence of the slab, it is convenient to introduce an additional figure of merit, $\chi_{id}$, as the ratio of the useful transmitted power $P_2^0$ to the power extracted by an ``ideal" load connected directly to the voltage source, i.e. $P_{id}=\frac{|\mathcal{E}|^2}{2R_2}$:
\be \chi_{id}=\frac{\omega^2|L_{21}|^2 R_2^2}{|Z_1^\eff Z_2|^2}\to \left(\frac{R_2}{R_2^\eff}\frac{\omega|L_{21}|}{|Z_1^\eff|}\right)^2,
\label{eq:chi_ideal}
\ee
where in the latter limit we again assumed $|Z_2|\to R_2^\eff$.
Note that both $\chi$ and $\chi_{id}$ may be greater than one, whereas $\eta\le 1$ by definition.

In order to maximize $\chi_{id}$, the source coil must operate at a frequency at which
$|Z_1^\eff|$ is minimal. By varying the capacitance $C_1$, one can eliminate the imaginary (reactive) part of $Z_1^\eff$; in that regime, the figure of merit (\ref{eq:chi_ideal}) becomes
\be \chi_{id}=\left(\frac{\omega|L_{21}|R_2}{R_1^\eff R_2^\eff}\right)^2.
\label{eq:chi_ideal_res}
\ee

We have expressed two useful figures of merit through the modified coil impedances, $Z_{mn}\equiv j \omega L_{mn}$, which are calculated in the subsequent sections. It is interesting to see how they relate to the shift in resonance frequency induced by the slab. The resonance frequencies are obtained by equating $Z_1^{\eff}=0$, which gives a polynomial equation for the frequency:
\be Z_1 Z_2 + \omega^2 L_{12}L_{21} =0.
\ee
Neglecting resistive loads ($R_1=R_2=0$), a biquadratic equation is obtained for the resonance frequency:
\be
\omega^4 \det(L_{mn})C_1C_2-\omega^2(L_{11}C_1+L_{22}C_2)+1=0.
\ee
Its solutions are
\be
\omega_{\pm}^2 = \frac{\omega_1^{-2}+\omega_2^{-2} \pm \sqrt{(\omega_1^{-2}-\omega_2^{-2})^2+4\Omega^{-4}}}
{2(\omega_1^{-2}\omega_2^{-2}-\Omega^{-4})},
\ee
where $\omega_m^2\equiv 1/(C_m L_{mm})$ and $\Omega^{-4}\equiv L_{12}L_{21}C_1C_2$. For the frequency difference we obtain
\be
\omega^2_{+}-\omega^2_{-}=\frac{\sqrt{(\omega_1^{-2}-\omega_2^{-2})^2+4\Omega^{-4}}}{\omega_1^{-2}\omega_2^{-2}-\Omega^{-4}}.
\ee
For a symmetric system ($\omega_1=\omega_2\equiv\omega_0$), the frequency shift to the lowest order in $\Omega^{-2}$ is
 $\Delta\omega\equiv \omega_+-\omega_- \approx \Omega^{-2}\omega_0^3$.
Since $\Omega^{-4}\propto L_{12}L_{21} \propto \chi \propto \chi_{id}$, we can see that in the small coupling ($\chi \ll 1$), lossless limit, a linear relationship exists between the power coupling efficiencies and $(\Delta\omega)^2$. This relationship is however linear only if the slab contributions to self-inductances $L_{11,22}$, and consequently to the resonance frequencies $\omega_{1,2}$, are disregarded.

\section{\label{sec:L12} Calculation of mutual inductance in the point-dipole approximation}

To describe the operation of current-carrying coils in the presence of a metamaterial slab, one needs to know the inductances $L_{mn}$.
Self-inductance of a wire loop is a function of the loop radius as well as the wire thickness.
On the other hand, mutual inductance $L_{12}$ can be determined with good accuracy from the value of magnetic flux density $\vec B$ generated by coil 1 evaluated at the center of coil 2. The distribution of $\vec B$ in space can be found in the point-dipole approximation, i.e. by neglecting the size of the source coil. In this section, we calculate $L_{12}$ assuming that both coils are negligibly small in comparison with the spatial scale of the magnetic field gradient. The problem is then reduced to the well-known problem of a dipole antenna over a flat layer of homogeneous medium. For an infinitely extended substrate, this problem was solved by Sommerfeld~\cite{sommerfeld_pde}, and the solution is often referred to as the {\it Sommerfeld integral}. For an infinitely wide slab of finite thickness, the solutions in terms of similar integrals were reported by Weng Chew~\cite{chew94}. Here, we generalize these solutions for a slab with uniaxially anisotropic properties. In the quasistatic limit, we are able to obtain closed-form expressions for all fields in terms of a well-known special function, Lerch transcendent $\Phi_L$.

We assume that a slab of thickness $D$, which occupies the space between $0\le z\le D$, is embedded into homogeneous medium, such as air, with relative permittivity $\epsilon_v$ and relative permeability $\mu_v$. The relative permittivity and permeability tensors of the slab are assumed uniaxial, with Cartesian components
\be \epsilon=\diag(\epsilon_x,\epsilon_y\equiv\epsilon_x,\epsilon_z), \mu=\diag(\mu_x,\mu_y\equiv\mu_x,\mu_z).\ee
Since the principal axis of the material properties is aligned with the normal to the slab,
the cylindrical symmetry of the problem is preserved, and one can use the approach of Chew~\cite{chew94} even with the medium anisotropy; the benefits of anisotropy are explained below. The source, a point magnetic dipole with magnetic moment $m_1=i_1A_1$, oriented parallel to the slab, is placed on the $z$-axis at $z=-d_1$.

On the reflection side of the slab, in the region with $-d_1<z<0$, the axial components of electric and magnetic field can be written as~\cite{chew94}
\ba
H_z=\frac{m_1}{8\pi}\cos\phi\int_{-\infty}^\infty d k_\rho k_\rho^2 H_1^{(1)}(k_\rho\rho)e^{jk_z d_1}
\left[e^{jk_z z}+R_{TE}(k_\rho)e^{-jk_z z}\right],\nonumber \\
E_z=\frac{\omega\mu_v\mu_0 m_1}{8\pi}\sin\phi\int_{-\infty}^\infty d k_\rho \frac{k_\rho^2}{k_z} H_1^{(1)}(k_\rho\rho)e^{jk_z d_1}
\left[e^{jk_z z}+R_{TM}(k_\rho)e^{-jk_z z}\right],
\label{eq:refl_Hz}
\ea
where $k_z=\sqrt{k_0^2\epsilon_v\mu_v-k_\rho^2}$, and $R_{TE}$ and $R_{TM}$ are reflection coefficients for the electric field in $TE$ and $TM$ polarizations, which are derived below.
On the transmission side ($z>D$), the fields are
\ba
H_z=\frac{m_1}{8\pi}\cos\phi\int_{-\infty}^\infty d k_\rho k_\rho^2 H_1^{(1)}(k_\rho\rho)T_{TE}(k_\rho)e^{jk_z(z-D)},\nonumber \\
E_z=\frac{\omega\mu_v\mu_0 m_1}{8\pi}\sin\phi\int_{-\infty}^\infty d k_\rho \frac{k_\rho^2}{k_z} H_1^{(1)}(k_\rho\rho)T_{TM}(k_\rho)e^{jk_z(z-D)}.
\label{eq:trans_Hz}
\ea

To facilitate the derivations, introduce the spectral components of all fields according to Ref.~\cite{chew94}
\ba
\vec H(\rho,\phi,z)= \int_{-\infty}^\infty d k_\rho \widetilde{\vec H}(k_\rho,z,\phi), \nonumber \\
\vec E(\rho,\phi,z)= \int_{-\infty}^\infty d k_\rho \widetilde{\vec E}(k_\rho,z,\phi),
\label{eq:spectral}
\ea
and notice that the problem reduces to the one-dimensional propagation of plane waves $\widetilde {\vec E}(k_\rho,z,\phi)$ in the $z$-direction through the layered medium. The reflection and transmission coefficients, calculated in accordance with our phase normalization choices seen from (\ref{eq:trans_Hz}), are calculated in the Appendix; see equations (\ref{eq:RT_TE},\ref{eq:RT_TM}).

The remaining four components of the $E$ and $H$ vector fields, are expressed through $\tilde H_z(k_\rho,z)$ and
$\tilde E_z(k_\rho,z)$ using the formulas of Chew~\cite{chew94} (in Cartesian coordinates):
\ba
\tilde H_s(k_\rho,\phi,z)= \frac{1}{k_\rho^2}\left[\nabla_s\frac{\d \tilde H_z}{\d z}+j\omega\epsilon_v\epsilon_0\hat z \times \nabla_s\tilde E_z \right],
\nonumber\\
\tilde E_s(k_\rho,\phi,z)= \frac{1}{k_\rho^2}\left[\nabla_s\frac{\d \tilde E_z}{\d z}-j\omega\mu_v\mu_0\hat z \times \nabla_s\tilde H_z \right],
\label{eq:transverse}
\ea
where subscript $s$ indicates the transverse part of a vector. In particular, for $\tilde H_x$ values on the axis, in the region $z>D$ (where the second coil would be placed), we obtain
\be
\tilde H_x(k_\rho,\rho=0,z)= \frac{j m_1}{16\pi}\left[k_\rho k_z T_{TE}-\frac{k_v^2k_\rho}{k_z}T_{TM} \right]e^{j k_z(z-D)},
\label{eq:Hx_axis}
\ee
where $k_v^2\equiv \epsilon_v\mu_v k_0^2$.
In deriving equation (\ref{eq:Hx_axis}), we have used $J_1'(0)=1/2$ and the oddness of the Bessel function $Y_1(k_\rho\rho)=\Im H_1^{(1)}(k_\rho\rho)$.

So far, the calculations have been exact. To simplify further analysis, assume that the distance between the coils is deeply sub-wavelength, i.e. $d\equiv D+d_1+d_2\ll \lambda_v=2\pi/k_v$. Since we are interested only in the near-field region around it, only Fourier components with $|k_\rho|\gg k_v$ are important, and the exact formulas above can be greatly simplified. Under this assumption, the following approximations can be made (as in Ref.~\cite{shvets_spie03}):
\ba \chi_0\equiv -j k_z \approx |k_\rho|, \\
\chi_{TE}\equiv \sqrt{k_\rho^2\mu_x/\mu_z-\mu_x \epsilon_x k_0^2}\approx |k_\rho|\sqrt{\mu_x/\mu_z}, \\
\chi_{TM}=\sqrt{k_\rho^2\epsilon_x/\epsilon_z-\epsilon_x \mu_x k_0^2}\approx |k_\rho|\sqrt{\epsilon_x/\epsilon_z}.
\ea
Note that we implicitly assume that $\sign\Re \mu_x=\sign\Re\mu_z$ and $\sign\Re \epsilon_x=\sign\Re\epsilon_z$;
in these regimes, spectral components evanescent in vacuum regions are also evanescent (or exponentially growing) inside the slab.
The reflection and transmission functions in this {\it quasistatic} approximation are
\ba
R_{TE}\approx \frac{g_{TE}\left(e^{\alpha_{TE} |k_\rho| D}-e^{-\alpha_{TE} |k_\rho| D}\right)}{a_{TE} e^{\alpha_{TE} |k_\rho| D}+b_{TE} e^{-\alpha_{TE} |k_\rho| D}},
\nonumber \\
T_{TE}\approx \frac{h_{TE} e^{-|k_\rho| d_1}}{a_{TE} e^{\alpha_{TE} |k_\rho| D}+b_{TE} e^{-\alpha_{TE} |k_\rho| D}},
\ea
where $\alpha_{TE}\equiv \sqrt{\mu_x/\mu_z}$, $a_{TE}\equiv -\left(\frac{\alpha_{TE}}{\mu_x}+\frac{1}{\mu_v}\right)^2$, $b_{TE}\equiv \left(\frac{\alpha_{TE}}{\mu_x}-\frac{1}{\mu_v}\right)^2$, $g_{TE}=\frac{1}{\mu_x\mu_z}-\frac{1}{\mu_v^2}$ and $h_{TE}=-\frac{4\alpha_{TE}}{\mu_x\mu_v}$ are material constants; analogous expressions for TM components are obtained by the substitution $\mu\to\epsilon$.

Magnetic field intensity at the position of the second dipole becomes
\ba
H_x(\rho=0,z=D+d_2)=\frac{-m_1}{16\pi}
\int_{-\infty}^\infty d k_\rho \left[k_\rho^2 T_{TE}(|k_\rho|)+k_v^2T_{TM}(|k_\rho|) \right]e^{-|k_\rho| d_2} \nonumber \\
=\frac{-m_1}{8\pi}\int_{0}^\infty d k_\rho \left[k_\rho^2 T_{TE}(k_\rho)+k_v^2T_{TM}(k_\rho) \right]e^{-k_\rho d_2}.
\label{eq:Hx_axis_approx}
\ea
Similarly, at the position of the first dipole we have
\ba
H_x(\rho=0,z=-d_1)
=\frac{-m_1}{8\pi}\int_{0}^\infty d k_\rho k_\rho^2 \left[ (1+R_{TE}(k_\rho)e^{-2k_\rho d_1})+k_v^2(1+R_{TM}(k_\rho)e^{-2k_\rho d_1}) \right].
\label{eq:Hx_refl}
\ea
The integral in (\ref{eq:Hx_refl}) diverges, as it includes the field generated by a point dipole.
The reflected portion of this field is nevertheless finite, and it will be used in the next section for a calculation of slab effect on self-impedance.

We observe that the contribution of the TM wave is suppressed by a very small factor, of order $(k_0D)^2$. Practically, it means that dielectric slabs cannot provide strong enhancement of the mutual inductance, thus the need for magnetic metamaterial slabs.

The only remaining integral in equation (\ref{eq:Hx_axis_approx}) can be reduced to the standard Lerch transcendent function ($\Phi_L$) using the substitution $t=\exp(-2\alpha D k_\rho)$. Noticing that
\be
\int_0^1 \frac{ (\ln t)^s t^{u-1}}{1+\frac{b}{a}t} dt = 2 \Phi_L(-\frac{b}{a},s+1,u),
\ee
we obtain the flux through coil 2:
\ba
\Phi_2=B_x(\rho=0,z=D+d_2)A_2=\frac{-m_1 \mu_v \mu_0 A_2}{8\pi}\int_{0}^\infty d k_\rho \left[k_\rho^2 T_{TE}+k_v^2 T_{TM}\right] e^{-k_\rho d_2} \nonumber \\
=\frac{-\mu_v \mu_0 i_1 A_1 A_2}{4\pi}\left[\frac{h_{TE}}{a_{TE}}\frac{1}{(2\alpha_{TE} D)^3}\Phi_L(-\frac{b_{TE}}{a_{TE}},3,u_{TE})+\frac{h_{TM}}{a_{TM}}\frac{k_v^2}{(2\alpha_{TM} D)}\Phi_L(-\frac{b_{TM}}{a_{TM}},1,u_{TM})\right],
\label{eq:Phi2_approx}
\ea
where $u_{TE,TM}=(\alpha_{TE,TM} D + d_1 + d_2)/(2\alpha_{TE,TM} D)>0$.
In the following, we will be neglecting the TM contribution, and omitting the TE subscript for all variables.
The mutual inductance coefficient is obtained from expression (\ref{eq:Phi2_approx}) by dividing out the current $i_1$:
\be
L_{21}=\Phi_2/i_1=\frac{-\mu_v \mu_0 A_1 A_2}{4\pi}\frac{h}{a(2\alpha D)^3}\Phi_L(-\frac{b}{a},3,u).
\label{eq:L21}
\ee

For positive real $u$, the function $\Phi_L(\gamma,3,u)$ is analytic in both $\gamma$ and $u$, except the line $\gamma=[1;+\infty)$ where it has a branch cut discontinuity in $\gamma$.
Since the branch cut discontinuity affects only the phase of $\Phi_L$, and the physical result depends only on $|\Phi_L|$, we can treat this function as continuous.

To highlight the effect of the slab on the inductive coupling, one should compare $L_{21}$ given by (\ref{eq:L21}) to its value in the absence of the slab, i.e. when $\mu_x=\mu_z=1$ and $\mu_v=1$. Substituting $a=h=-4$ and $b=g=0$ into (\ref{eq:L21}) and using $\Phi_L(0,3,u)=u^{-3}$, we obtain
\be
L_{21}^{vac}=
\frac{-\mu_v \mu_0 A_1 A_2}{4\pi}\frac{1}{(2 D)^3}\Phi_L(0,3,d/(2 D))\equiv
\frac{-\mu_v \mu_0 A_1 A_2}{4\pi d^3}.
\label{eq:L21_vac}
\ee
where $d=D+d_1+d_2$ is the distance between the coils.

\section{\label{sec:L11} The effect of the slab on self-inductance}

The contribution of the slab response to the self-inductance of coil 1 can be also estimated in the point-dipole approximation using the following model~\cite{dong_chan10}. Consider a point magnetic dipole $\vec m_1=m_1 \hat x$  excited by the sum of external magnetic field, $H_x^{ext}$, and the field reflected from the slab,
\be H_x^{ref}=G_{xx}^{ref}m_1,
\label{eq:Hx_ref_G}
\ee
where
\be G_{xx}^{ref}=\frac{1}{8\pi}\int_{0}^\infty d k_\rho e^{-2k_\rho d_1} \left[k_\rho^2 R_{TE}(k_\rho)-k_0^2 R_{TM}(k_\rho) \right]
\ee
is the reflected portion of the Green's function, as seen from equation (\ref{eq:Hx_refl}). The total magnetic moment equals
\be m_1=\alpha_m(H_x^{ext}+H_x^{ref}),
\label{eq:m1_total}
 \ee
where $\alpha_m$ is the {\it intrinsic} magnetic polarizability of the dipole unperturbed by the slab.
Resolving equations (\ref{eq:Hx_ref_G},\ref{eq:m1_total}), we find $m_1=\alpha_m/(1-\alpha_m G_{xx}^{ref})H_x^{ext}$, which means that the effective polarizability of the dipole in the presence of the slab is
\be
\alpha_m^\eff=(\alpha_m^{-1}-G_{xx}^{ref})^{-1}.
\label{eq:alpha_m_eff}
\ee
From the relationship $\Phi_1=L_{11}i_1=L_{11}m_1/A_1$ we find that this polarizability relates to self-inductance as $L_{11}=\mu_v \mu_0 A_1^2/\alpha_m$; thus,
\be
L_{11}=L_{11}^{(0)}-\mu_v \mu_0A_1^2G_{xx}^{ref},
\label{eq:L11_eff}
\ee
where $L_{11}^{(0)}$ is the self-inductance of coil 1 without the slab.
In the quasistatic approximation, and neglecting the weak contribution of the TM wave, we obtain the slab contribution to self-inductance $L_{11}$:
\be L_{11}^{(1)}=-\mu_v\mu_0A_1^2 G_{xx}^{ref}\approx \frac{\mu_v\mu_0A_1^2}
{4\pi(2\alpha D)^3}\frac{g}{a}\left[\Phi_L(-\frac{b}{a},3,v)-\Phi_L(-\frac{b}{a},3,v+1)\right],
\label{eq:L11_approx}
\ee
where $v=d_1/(\alpha D)$. Using the identity
\be \Phi_L(\gamma,s,u+1)=\frac{1}{\gamma}\left(\Phi_L(\gamma,s,u)-u^{-s}\right),
\ee
expression (\ref{eq:L11_approx}) for $v>0$ can be reduced to
\be L_{11}^{(1)}= \frac{\mu_v\mu_0A_1^2}
{4\pi(2\alpha D)^3}\frac{g}{a}\left[\left(1+\frac{a}{b}\right)\Phi_L(-\frac{b}{a},3,v)-\frac{a}{b}v^{-3}\right].
\label{eq:L11_reduced}
\ee
The expression for $L_{22}$ is obtained by replacing $v=d_1/(\alpha D)$ in (\ref{eq:L11_reduced}) with $w=d_2/(\alpha D)$.

\section{\label{sec:transfer} Power transfer efficiency analysis}

Now we turn to the most realistic scenario where the self-resonant coils are situated in free space ($\epsilon_v=\mu_v=1$).
Optimization of the power transfer efficiency,
\be
\eta = \frac{R_2}{R_2-\omega \Im[\frac{g}{a}\Phi_L(-b/a,3,w)]}\frac{\chi}{1+\chi},
\ee
where
\be
\chi \approx
\omega^2\left(\frac{h}{a}\right)^2
\frac{[\Phi_L(-b/a,3,u)]^2}{\left( R_1-\omega\Im [\frac{g}{a}\Phi_L(-b/a,3,v)]\right)\left(R_2-\omega \Im[\frac{g}{a}\Phi_L(-b/a,3,w)]\right)}
,
\label{eq:chi_final}
\ee
requires finding the optimum balance between the enhancement of $L_{12}$ and the growth of $\Im L_{11,22}^{(1)}$; it is easy to see that increasing the former also leads to the increase of the latter.
In expression (\ref{eq:chi_final}), the term $\Phi_L(-b/a,3,v+1)$ from (\ref{eq:L11_approx}) was neglected in comparison with $\Phi_L(-b/a,3,v)$. This is a valid approximation for small $|a|$ and $0\le v \le 1$; the reason why we are only interested in this regime is explained below.

For $0\le u \le 1$, $|\frac{1}{a}\Phi_L(-b/a,3,u)|$ as a function of the complex parameter $a$ has a sharp peak at $a=0$.
Therefore, we may expect that the maximum of $\chi$ (and $\eta$) is obtained at the minimum of $|a|\equiv \left|\frac{\alpha}{\mu_x}+\frac{1}{\mu_v}\right|^2$. For a lossless magnetic metamaterial slab, $|a|$ can be precisely zero when the {\it perfect lens} condition
 \be \mu_x=-\alpha \mu_v
 \ee
is met. In the point-dipole approximation, a lossless magnetic metamaterial slab can thus provide an {\it infinite} mutual inductance, leading to the perfect power transfer efficiency, $\eta=1$.

In a real system, $\eta$ is limited by losses; its maximum is achieved by minimizing $|a|$, i.e.
by requiring
\be
\Re\mu_x=-\Re(\alpha) \mu_v.
\label{eq:magn_superlensing}
\ee
In this regime, a closed-form analytic expression for the mutual inductance can be obtained in the limit of small losses. Suppose that the magnetic loss tangents $\sigma_{x,z}$, introduced according to $\mu_{x,z}=\mu_{x,z}^r(1+j \sigma_{x,z})$, are both small, and $\mu_x^r=-\mu_v\sqrt{\mu_x^r/\mu_z^r}$. Note that we assume that both $\mu_x^r$ and $\mu_z^r$ are negative.
Then $a\approx \sigma^2/\mu_v^2$, where $\sigma\equiv(\sigma_x+\sigma_z)/2$. The other parameters in the expression (\ref{eq:chi_final}) become approximately (to the $O(\sigma)$ accuracy):
\ba
b=4/\mu_v^2,\\
h=4/\mu_v^2, \\
g=\frac{-2j\sigma}{\mu_v^2}.
\ea
To the same accuracy, the expression (\ref{eq:chi_final}) becomes
\ba
\eta = \frac{R_2/\tilde Z_2}{R_2/\tilde Z_2+\frac{1}{\sigma}\Phi_L(-b/a,3,w)}\frac{\chi}{1+\chi},\nonumber\\
\chi = \frac{\frac{4}{\sigma^4}\Phi_L^2(-4/\sigma^2,3,(1+v+w)/2)}{\left(\frac{R_1}{\tilde Z_1}+\frac{1}{\sigma}\Phi_L(-4/\sigma^2,3,v)\right)
\left(\frac{R_2}{\tilde Z_2}+\frac{1}{\sigma} \Phi_L(-4/\sigma^2,3,w)\right)},
\label{eq:chi_lowloss}
\ea
where we have introduced two parameters with the dimensions of impedance:
\ba
\tilde Z_{1,2}\equiv Z_0 \frac{A_{1,2}^2}{\lambda(2\alpha D)^3}.
\ea
Here, $\lambda\equiv 2\pi c/\omega$ is the vacuum wavelength and $Z_0\equiv \mu_0 c$ is the free-space impedance.

Before we find the optimum coupling regime, we can make two interesting observations about the behavior of Lerch function at large negative values of its first argument. First, one can show that
\ba
\lim_{a\to 0} \frac{1}{a}\Phi_L(-b/a,3,u)=
\left[
\begin{array}{l l}
\infty, & 0\le u\le 1, \\
\frac{2}{b(u-1)^3}<\infty, & u>1.
\end{array}
\right.
\ea
Therefore, with arbitrarily small loss tangents, arbitrarily large enhancement of the mutual inductance $L_{12}$ can be obtained when $0\le u \le 1$, i.e. $d_1+d_2\le \alpha D$.


Second, the slab contribution to the coil self-impedance
\be
L_{11}^{(1)} \propto \frac{1}{\sigma}\Phi_L(-4/\sigma^2,3,v)
\ee
diverges in the limit $\sigma\to 0$ whenever $v\le 1/2$, i.e. when $d_1<\alpha D/2$, but converges to a finite limit for any $v>1/2$, in agreement with the findings of Refs.~\cite{milton_rsoca06,dong_chan10}. This means that the dipole source can be completely {\it quenched}~\cite{dulkeith_gittins_prl02} by a lossless negative-$\mu$ slab in the superlensing regime (\ref{eq:magn_superlensing}) whenever the source is closer to the slab than the {\it quenching distance}, $d_q=\alpha D/2$. However, one should realize that with a finite loss,
$L_{11}$ remains finite for any $d_1\ge 0$; total quenching is possible only with a non-physical choice $\sigma=0$.

\begin{figure}
\centering
\begin{tabular}{c c c}
\includegraphics[width=0.3\columnwidth]{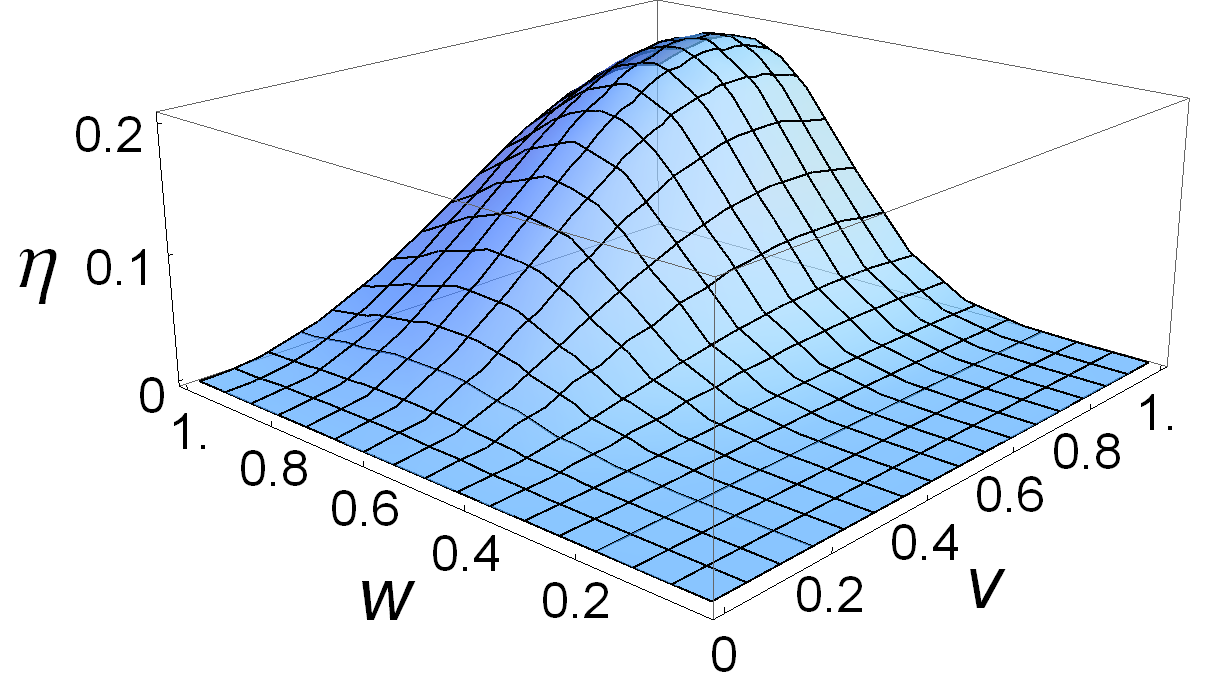}&
\includegraphics[width=0.3\columnwidth]{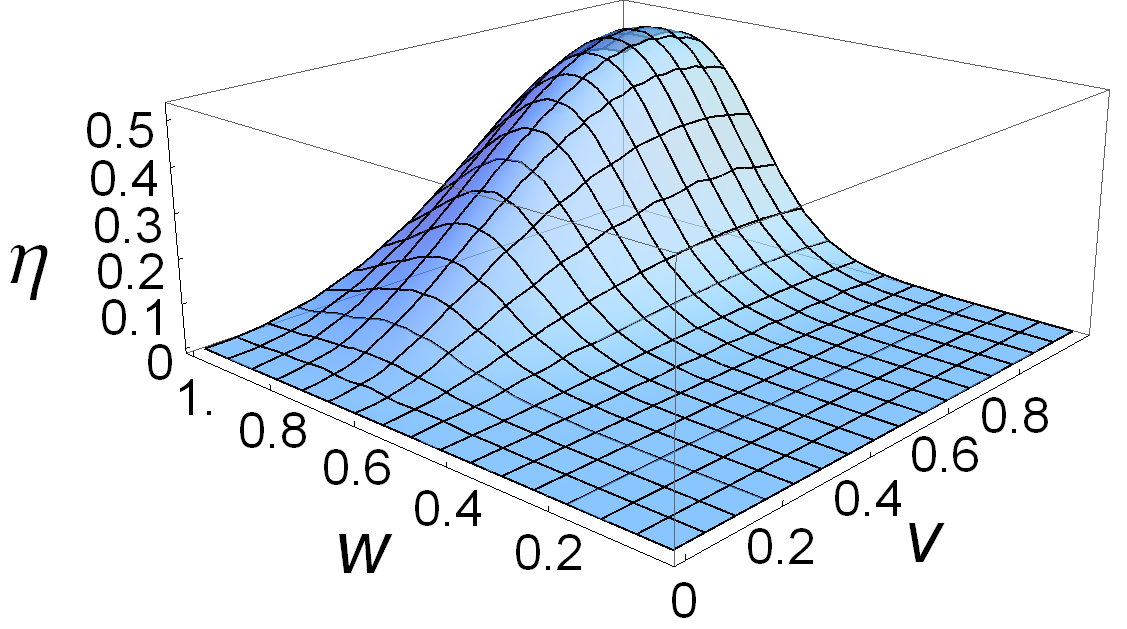}&
\includegraphics[width=0.3\columnwidth]{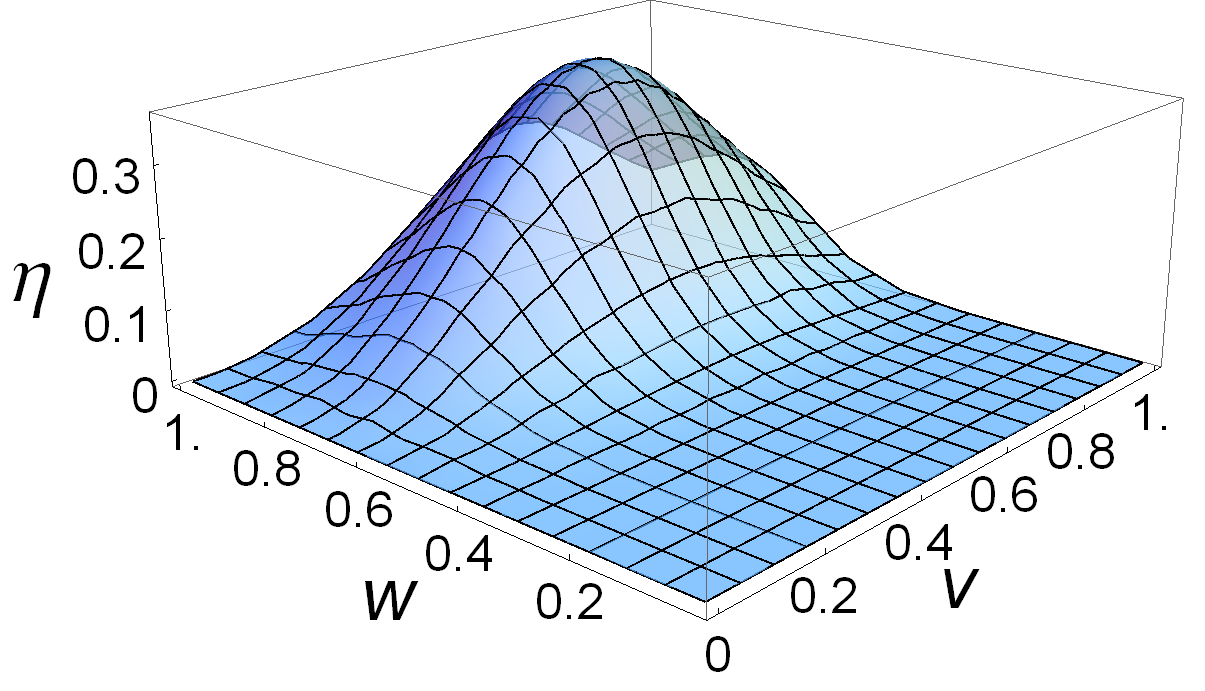}
\\
(a)&(b)&(c)\\
\end{tabular}
\caption{
(Color online)
Power transfer efficiency $\eta$ as a function of the normalized distances $v$ and $w$.
(a) $R_1/\tilde Z_1=1$, $R_2/\tilde Z_2=10$, $\sigma=0.1$.
(b) Same as (a) but $\sigma=0.01$.
(c) $R_1/\tilde Z_1=10$, $R_2/\tilde Z_2=10$, $\sigma=0.01$.
}
\label{fig:eta_vs_vw}
\end{figure}

\begin{figure}
\centering
\begin{tabular}{cc}
\includegraphics[width=0.45\columnwidth]{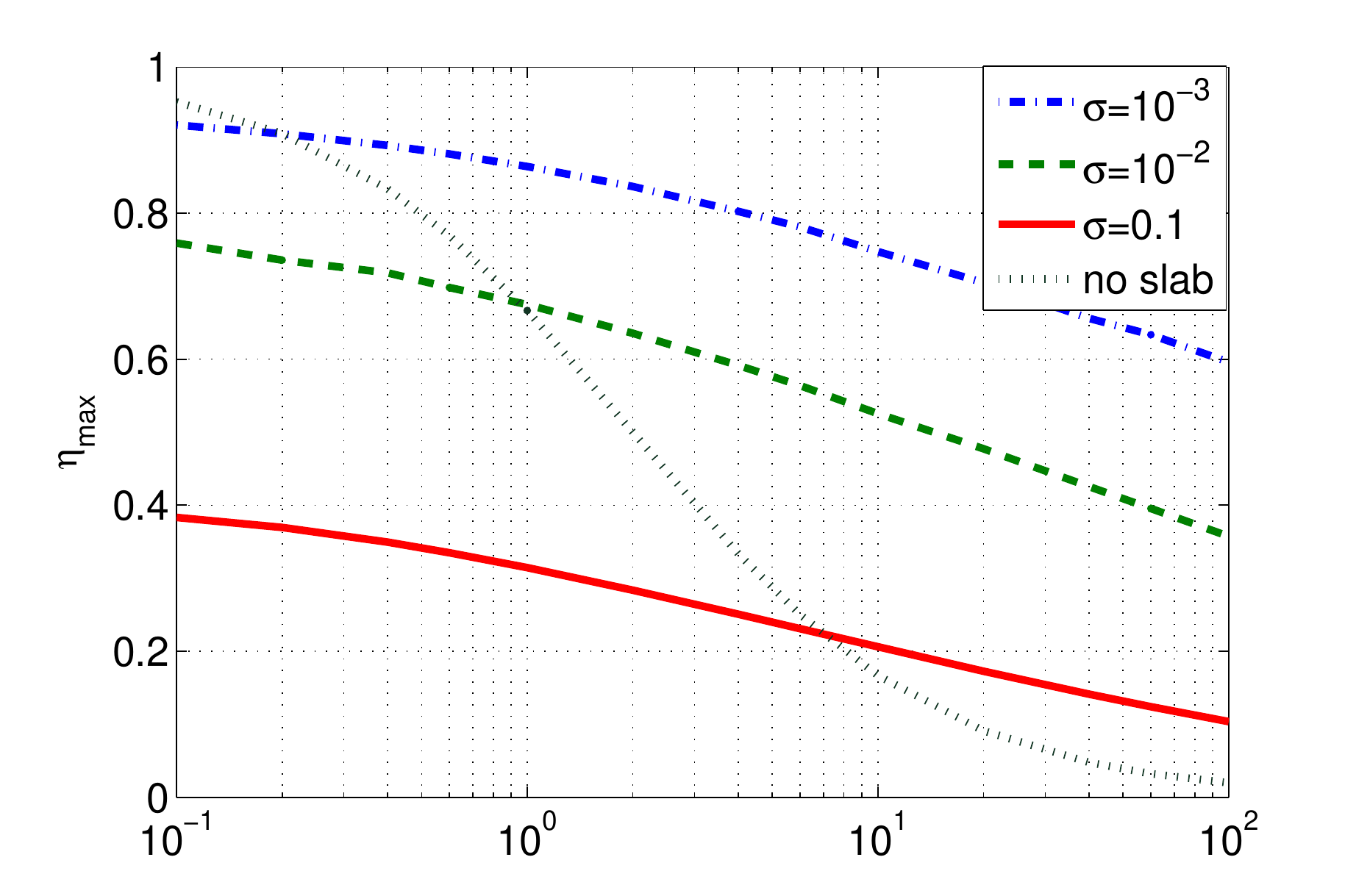}&
\includegraphics[width=0.45\columnwidth]{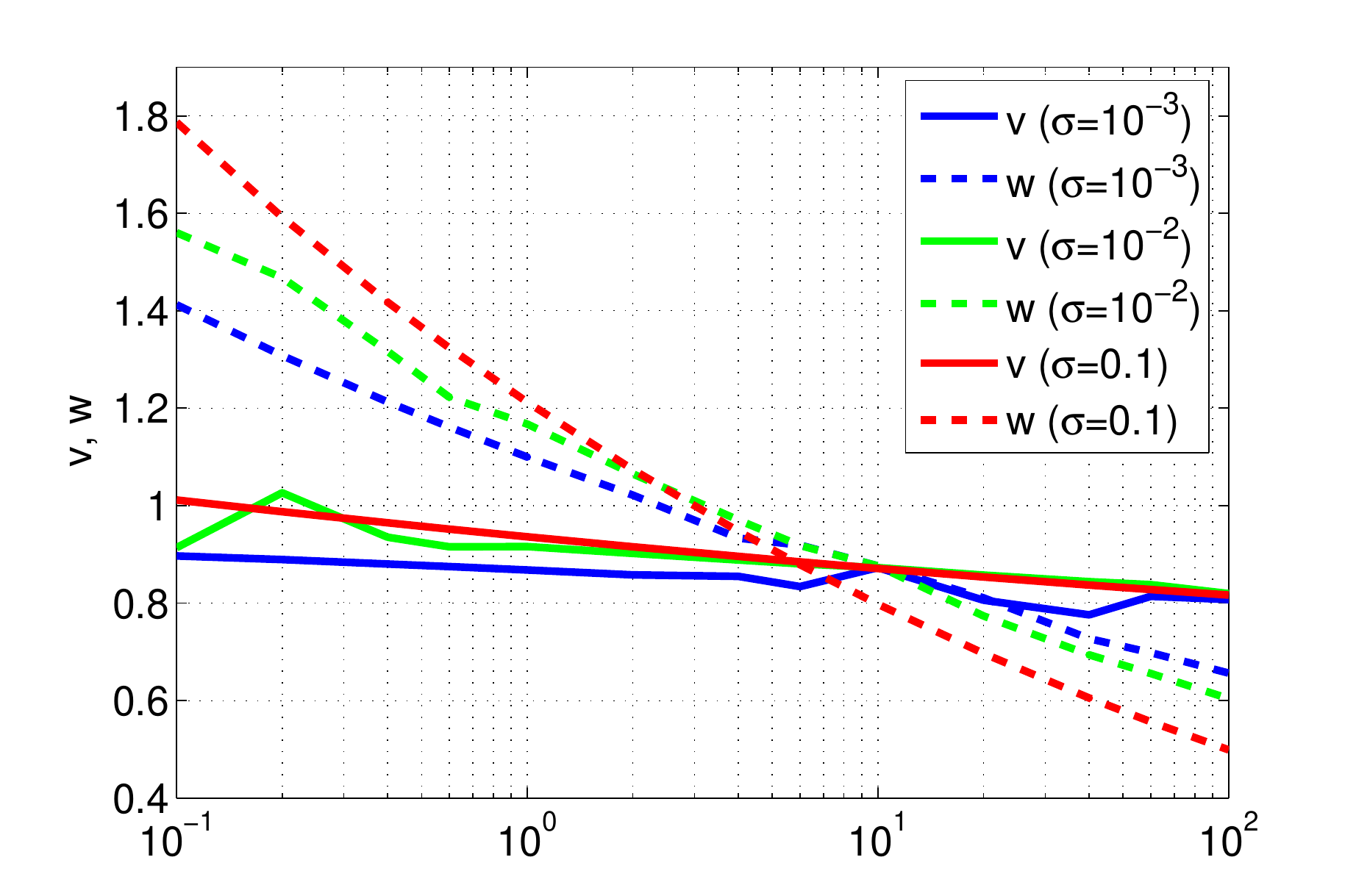}\\
(a)&(b)\\
\end{tabular}
\caption{
(Color online)
Left: Power transfer efficiency $\eta$, maximized over normalized distances $(v,w)$, as a function of the normalized resistance $R_2/\tilde Z_2$ for $R_1/\tilde Z_1=1$. For comparison, power transfer efficiency in free space -- $\eta^{vac}$ from equation (\ref{eq:chi_vac} -- is also shown as the black dotted curve.
Right: The optimum normalized distances $v$ (solid curves) and $w$ (dashed curves) in the same configuration. Curve coloring corresponds to loss tangents $\sigma=0.001$ (blue), $0.01$ (green), and $0.1$ (red).
}
\label{fig:maxeta_vs_R}
\end{figure}

The function $\eta(v,w)$ has a well-defined maximum, defining the optimum positions of the coils relative to the slab, as seen in Figure~\ref{fig:eta_vs_vw}. The optimum positions depend on the resistances $R_{1,2}$; a larger resistance $R_1$ ($R_2$) requires coil 1 (coil 2) to be closer to the slab's left (right) interface in order to achieve maximum performance.

The maximum value of the function $\eta$, maximized over the distance variables $(v,w)$, is a monotonically decreasing function of the loads $R_1$ and $R_2$. It can be computed numerically for any given loss tangent $\sigma$. Figure~\ref{fig:maxeta_vs_R}(a) shows the maximum power efficiency for several choices of $\sigma$. The optimum positions of the coils, $v_{\max}$ and $w_{\max}$, are shown in Figure~\ref{fig:maxeta_vs_R}(b).

To find the regime in which a metamaterial slab offers significant improvement for the power transfer system, compare our result (\ref{eq:chi_lowloss}) with the no-slab case, for which
\ba
\eta=\eta^{vac}\equiv \frac{\chi^{vac}}{1+\chi^{vac}},\nonumber \\
\chi=\chi^{vac}\equiv 2\frac{\tilde Z_1^{vac}}{R_1}\frac{\tilde Z_2^{vac}}{R_2},
\label{eq:chi_vac}
\ea
where $\tilde Z_{1,2}^{vac}\equiv \tilde Z_{1,2}|_{\alpha D=d}$.
The expression (\ref{eq:chi_vac}) is small whenever either or both of the circuits are loaded with resistances exceeding the reference impedances $\tilde Z_{1,2}^{vac}$. For example, if both $R_{1,2}/\tilde Z_{1,2}^{vac}$ ratios are of order $10$, $\chi^{vac}$ is of order $10^{-2}$. For the same parameters, however, $\chi$ given by expression (\ref{eq:chi_lowloss}) is of order $0.1-0.3$ even with realistic loss tangents in the $0.1-0.01$ range. In this regime, the slab offers at least one order of magnitude improvement of the power transfer efficiency relative to the no-slab case. For even smaller loss tangents, efficiency $\eta$ from expression (\ref{eq:chi_lowloss}) can be close to $100\%$, as seen from Figure~\ref{fig:maxeta_vs_R}(a). As expected, $\eta$ decreases monotonically with increasing loss tangent; it diverges slowly (logarithmically) in the limit $\sigma\to 0$.

\section{\label{sec:orientation} The effect of dipole orientation}

So far, we have considered one orientation of the dipoles, namely, two dipoles parallel to the slab and to each other.
Our formalism lets us consider two additional configurations: (a) two dipoles polarized in the $z$-direction, and (b) one $x$-oriented and one $z$-oriented dipole. To consider these cases, our previous derivations must be complemented with a calculation of magnetic field of the $z$-oriented dipole. Following Chew~\cite{chew94}, the field of a magnetic dipole with moment $\vec m_1=m_1 \hat z$ pointing in the positive $z$-direction is written as
\ba
H_z(\rho, \phi, z)=\frac{j m_1}{8\pi}\int_{-\infty}^\infty dk_\rho \frac{k_\rho^3}{k_z}H_0^{(1)}(k_\rho\rho)T_{TE}e^{jk_z(z-D)}, \\
H_z(\rho, \phi, z)=\frac{j m_1}{8\pi}\int_{-\infty}^\infty dk_\rho \frac{k_\rho^3}{k_z}H_0^{(1)}(k_\rho\rho)(1+R_{TE}e^{-jk_z z})e^{j k_z d_1},
\ea
in the transmitted ($z>D$) and reflected ($-d_1<z<0$) regions, respectively.
For the magnetic field at the locations of the two dipoles we obtain
\ba
H^{trans}_z(\rho=0,z=D+d_2)=\frac{j m_1}{4\pi}\int_{0}^\infty dk_\rho \frac{k_\rho^3}{k_z}T_{TE}e^{j k_z d_2}, \\
H^{ref}_z(\rho=0,z=-d_1)=\frac{j m_1}{4\pi}\int_{0}^\infty dk_\rho \frac{k_\rho^3}{k_z}R_{TE}e^{2 j k_z d_1}.
\ea
These fields are different from the expressions (\ref{eq:Hx_axis_approx},\ref{eq:Hx_refl}) for the $m_x-m_x$ case by a factor of two, but are otherwise the same. This means that $L_{11}^{(1)}$ and $L_{21}=L_{12}$ for the $m_z-m_z$ case are twice the corresponding inductances in the $m_x-m_x$ case. The coupling efficiency formula (\ref{eq:chi_lowloss}) suggests that $\chi$ for the $m_z-m_z$ case with load resistances $R_{1,2}$ is the same as $\chi$ for the $m_x-m_x$ case with resistances $R_{1,2}'=R_{1,2}/2$. Since the coupling efficiency increases with decreasing loads, the $m_z-m_z$ polarization provides a somewhat better coupling, with all other parameters fixed. The exact amount of additional enhancement can be determined from the diagram shown in Figure~\ref{fig:maxeta_vs_R}(a).


Finally, we note that in our configuration with the two dipoles aligned to the $z$-axis, there is no coupling between $m_x$ and $m_z$ components: the axial magnetic field $H_z$ of the $x$-polarized dipoles vanishes on the $z$-axis.

\section{\label{sec:dielectric} On the feasibility of efficient coupling of magnetic dipoles through a dielectric-only slab}

So far, we have been assuming that the coupling between the magnetic dipoles is mediated by the TE waves, which can be greatly enhanced by a negative-$\mu$ slab; the contribution of the TM wave was neglected.
Considering that metamaterials with electric-only response may offer lower loss tangents, it is interesting to see if a negative-$\epsilon$ slab can compete with a negative-$\mu$ slab.

Consider a situation where only the TM waves are enhanced, such that the TM terms in expressions (\ref{eq:Hx_axis_approx}) and (\ref{eq:Hx_refl}) dominate. This regime is achieved by choosing a dielectric-only slab with unit relative permeability and anisotropic relative permittivity $\epsilon_{x,z}<0$, and then tuning the real parts of $\epsilon_{x,z}$ to the superlensing condition, $\Re \epsilon_x=-\Re(\alpha_{TM})\epsilon_v$.
The mutual inductance (\ref{eq:Phi2_approx}) in this regime becomes
\be
L_{21}=\frac{\mu_0 A_1A_2}{4\pi}\left[ \frac{1}{d^3}+\frac{4k_0^2}{\sigma_e^2(2\alpha_{TM}D)}\Phi_L(-\frac{4}{\sigma_e^2},1,u_{TM})\right],
\label{L21_diel}
\ee
where $\sigma_e=(\sigma_{e,x}+\sigma_{e,z})/2$ is the effective dielectric loss tangent.
The first term in (\ref{L21_diel}) represents TE-wave mediated magnetic field transmission in free space.
The second term in (\ref{L21_diel}) dominates over the former whenever
\be
\frac{2}{\sigma_e^2}\Phi_L(-\frac{4}{\sigma_e^2},1,u_{TM})\gg \left(\frac{\lambda}{2\pi d}\right)^2.
\ee
Mathematically, it is possible to find a sufficiently small $\sigma_e$ for any $0<u_{TM}<1$ to satisfy this inequality, even with arbitrarily large $\lambda/d$ ratio.
The slab contribution to self-inductance is then dominated by the TM waves:
\be L_{11}^{(1)}= \frac{\mu_0A_1^2k_0^2}
{4\pi(2\alpha_{TM} D)}\frac{-2j}{\sigma_e}\Phi_L(-\frac{4}{\sigma_e^2},1,v_{TM}).
\label{eq:L11_TM}
\ee

In this regime, power transfer efficiency looks formally the same as the equation (\ref{eq:chi_lowloss}), in which all quantities are replaced with their TM analogs, except that the {\it characteristic} impedances $\tilde Z_{1,2}$ are now equal to
\be \tilde Z_{1,2}^{e} = Z_0 \frac{A_{1,2}^2 k_0^2}{\lambda(2\alpha_{TM}D)} = Z_0 \frac{A_{1,2}^2 (2\pi)^2}{\lambda^3(2\alpha_{TM}D)}.
\ee
Since the characteristic impedances $\tilde Z_{1,2}^{e}$ set the scale for attainable load resistances $R_{1,2}$, the dielectric slab can support only much smaller loads in the regime with reasonably high power transfer efficiency, $\eta>0.1$, in comparison with the magnetic slab. To achieve the same coupling efficiency with the same loads $R_{1,2}$ that the negative-$\mu$ slab can sustain, the loss tangent $\sigma_e$ must be decreased to an exponentially small value, which is not feasible in practice.
This finding is not applicable to two-dimensional (infinitely long) systems, where efficient coupling of two-dimensional point magnetic dipoles mediated by TM waves can be achieved with negative-$\epsilon$ slabs~\cite{shvets_spie03}.

\section{\label{sec:conclusion} Conclusions}

The rigorous analysis presented in sections \ref{sec:circuit}-\ref{sec:transfer} enables us to make the following conclusions about the metamaterial-based power relay system.

First, for given frequency $\omega$, metamaterial slab thickness $D$, magnetic anisotropy ratio $\alpha_{TE}$, and source and load resistances $R_{1,2}$, there exists a global maximum of the power transfer efficiency $\eta$ as a function of metamaterial permeability $\mu_z$ and the positions of the coils $d_{1,2}$ relative to the slab. It is also possible to show that a global optimum exists with respect to variable parameters $\mu_z$, $\alpha_{TE}$ and $D$ when $\omega$, $R_{1,2}$ and the total distance between the coils $d=d_1+d_2+D$ are fixed.

The two parameters -- slab thickness $D$ and anisotropy ratio $\alpha_{TE}$ -- can leverage each other~\cite{schurig_smith_njp05,gallina_engheta10}. Electromagnetically, the TE wave properties of an anisotropic slab of thickness $D$ and anisotropy $\alpha_{TE}$ are indistinguishable from those of an isotropic slab of thickness $D'=D\alpha_{TE}$. This statement can be derived using the invariance of Maxwell's equations with respect to coordinate transformations, a concept dubbed {\it transformation optics} in the recent years~\cite{schurig_smith06}. In practice, this means that the thickness of a metamaterial slab can be reduced without affecting performance, assuming that the magnetic loss tangent $\sigma$ can be kept constant while increasing the magnitude of the negative real part of $\mu_x$.

As a function of the useful load ($R_2$), optimum power transfer efficiency decreases slowly, much slower than the power law $1/R_2$ pertaining to the no-slab (vacuum) case. Consequently, there exists a threshold resistance $R_t$ such that the metamaterial system loaded with $R_2>R_t$ performs better than no-slab system. This threshold resistance $R_t$ decreases with the metamaterial loss tangent $\sigma$: less lossy metamaterials ``beat" free space performance at smaller loads.

Similar statements can be made about the efficiency as a function of the distance between the dipoles, $d$. Assuming that the anisotropy ratio is fixed, optimization of $\eta$ gives certain fixed ratios of the distances $d_{1,2}$ to the slab thickness $D$, and thus also the optimized ratio $D/d$. Since the power transfer efficiency depends on $D$ (and, after optimization, indirectly on $d$) only through the ratios $R_{1,2}/\tilde Z_{1,2}$, where $\tilde Z_{1,2}\propto 1/D^3 \propto 1/d^3$, Figure~\ref{fig:eta_vs_vw} can be used to determine how much $\eta$ decreases with increasing $D$. In the no-slab case, $\eta$ falls off with distance very quickly, as $1/d^{6}$. Again, the advantage of the metamaterial slab here is that it extends the range of distances $d$ within which acceptable efficiency can be obtained.

As typical for metamaterial superlenses, the performance can be theoretically arbitrarily good -- in our case, the power transfer efficiency can be as close to $100\%$ as needed -- but only if arbitrarily small loss tangents can be implemented. The growth of the figure of merit with decreasing loss is very slow (typically, logarithmic). Nevertheless, even with practically attainable loss tangents $\sigma\sim 0.1$ the metamaterial relay system can over-perform free space coupling efficiency by an order of magnitude or more, depending on the load resistance.

\section*{\label{sec:ack} Acknowledgements}

The authors are thankful to Da Huang (Duke University), Bingnan Wang and Koon Hoo Teo (Mitsubishi Electric Research Laboratories) for stimulating discussions relating to power transfer with magnetic metamaterials.

\section*{\label{sec:appendix} Appendix: Calculation of reflection and transmission coefficients for a slab}

Consider a slab of homogeneous, uniaxially anisotropic medium occupying the space between $0<z<D$, and a point magnetic dipole on the $z$-axis at $z=-d_1$, as shown in Fig.\ref{fig:coils_slab}.
The spectral components of electric and magnetic fields can be derived from the $H_z$ and $E_z$ fields of the TE and TM waves, whose evolution along the $z$-axis can be described as follows:

\ba
\tilde H_z=\frac{-m_1}{8\pi}\cos\phi k_\rho^2 H_1^{(1)}(k_\rho\rho)
 \times\left\{
 \begin{array}{l l}
 e^{-\chi_0 d_1}\left[e^{-\chi_0 z} + R_{TE}(k_\rho)e^{\chi_0 z}\right],& -d_1<z<0, \\
 t_1 e^{-\chi_{TE} z} + r_1(k_\rho)e^{\chi_{TE}z},& 0<z<D, \\
 T_{TE}e^{-\chi_0 (z-D)},& z>D, \\
 \end{array}
 \right.
 \label{eq:Hz_evol}
\ea

\ba
\tilde E_z=\frac{\omega\mu_v \mu_0 m_1}{8\pi}\sin\phi \frac{k_\rho^2}{k_z} H_1^{(1)}(k_\rho\rho)
\times\left\{
 \begin{array}{l l}
 e^{-\chi_0 d_1}\left[e^{-\chi_0 z} + R_{TM}(k_\rho)e^{\chi_0 z}\right],& -d_1<z<0, \\
 p_1 e^{-\chi_{TM} z} + q_1(k_\rho)e^{\chi_{TM}z},& 0<z<D, \\
 T_{TM}e^{-\chi_0 (z-D)},& z>D. \\
 \end{array}
 \right.
 \label{eq:Ez_evol}
\ea

In the above expressions,
\ba
\chi_0=\sqrt{k_\rho^2-\epsilon_v\mu_v k_0^2}=-j k_z, \\
\chi_{TE}=\sqrt{k_\rho^2\mu_x/\mu_z-\mu_x \epsilon_x k_0^2}, \\
\chi_{TM}=\sqrt{k_\rho^2\epsilon_x/\epsilon_z-\epsilon_x \mu_x k_0^2}.
\ea

From the continuity of $\tilde E_y=-\frac{j\omega \mu_x}{k_\rho^2} \frac{\d \tilde H_z}{\d x}$ and
$\tilde H_x=-\frac{1}{j\omega \mu_x}\frac{\d\tilde E_y}{\d z}=\frac{1}{k_\rho^2} \frac{\d^2\tilde H_z}{\d x\d z}$, it follows that
$\mu_x\tilde H_z$
and $\frac{\d \tilde H_z}{\d z}$ are continuous. From these requirements
and equations (\ref{eq:Hz_evol}), one obtains four equations for the four unknowns $R_{TE}, T_{TE}, t_1, r_1$:
\ba
e^{-\chi_0 d_1}\mu_v(1+R_{TE})=\mu_x(t_1+r_1),\\
\mu_x(t_1 e^{-\chi_{TE}D}+r_1 e^{\chi_{TE}D})=\mu_v T_{TE},\\
e^{-\chi_0 d_1}\chi_0(-1+R_{TE})=\chi_{TE}(-t_1+r_1),\\
\chi_{TE}(-t_1 e^{-\chi_{TE}D}+r_1 e^{\chi_{TE}D})=-\chi_0 T_{TE}.
\label{eq:TE_rt}
\ea

The relevant solutions are

\ba
R_{TE}=-\frac{\left(\frac{\chi_{TE}^2}{\mu_x^2}-\frac{\chi_0^2}{\mu_v^2}\right)\left[\exp(\chi_{TE} D)-\exp(-\chi_{TE} D)\right]}
{
\left(\frac{\chi_{TE}}{\mu_x}+\frac{\chi_0}{\mu_v}\right)^2 \exp({\chi_{TE} D})-\left(\frac{\chi_{TE}}{\mu_x}-\frac{\chi_0}{\mu_v}\right)^2 \exp({-\chi_{TE} D})},
\nonumber \\
T_{TE}=\frac{4\frac{\chi_{TE}}{\mu_x}\frac{\chi_0}{\mu_v} \exp(-\chi_0 d_1)}
{\left(\frac{\chi_{TE}}{\mu_x}+\frac{\chi_0}{\mu_v}\right)^2 \exp({\chi_{TE} D})-\left(\frac{\chi_{TE}}{\mu_x}-\frac{\chi_0}{\mu_v}\right)^2 \exp({-\chi_{TE} D})}.
\label{eq:RT_TE}
\ea

Similarly, for TM polarization we use continuity of $\epsilon_x \tilde E_z$ and $\frac{\d \tilde E_z}{\d z}$ with equations (\ref{eq:Ez_evol}) to yield
\ba
e^{-\chi_0 d_1}\epsilon_v(1+R_{TM})=\epsilon_x(p_1+q_1),\\
\epsilon_x(p_1 e^{-\chi_{TM}D}+q_1 e^{\chi_{TM}D})=\epsilon_v T_{TM},\\
e^{-\chi_0 d_1}\chi_0(-1+R_{TM})=\chi_{TM}(-p_1+q_1),\\
\chi_{TM}(-p_1 e^{-\chi_{TM}D}+q_1 e^{\chi_{TM}D})=-\chi_0 T_{TM}.
\label{eq:TM_rt}
\ea

The coefficients of interest are

\ba
R_{TM}=-\frac{\left(\frac{\chi_{TM}^2}{\epsilon_x^2}-\frac{\chi_0^2}{\epsilon_v^2}\right)\left[\exp({\chi_{TM} D})-\exp({-\chi_{TM} D})\right]}
{
\left(\frac{\chi_{TM}}{\epsilon_x}+\frac{\chi_0}{\epsilon_v}\right)^2 \exp({\chi_{TM} D})-
\left(\frac{\chi_{TM}}{\epsilon_x}-\frac{\chi_0}{\epsilon_v}\right)^2 \exp({-\chi_{TM} D})},
\nonumber \\
T_{TM}=\frac{4\frac{\chi_{TM}}{\epsilon_x}\frac{\chi_0}{\epsilon_v} \exp(-\chi_0 d_1)}
{\left(\frac{\chi_{TM}}{\epsilon_x}+\frac{\chi_0}{\epsilon_v}\right)^2 \exp({\chi_{TM} D})
-\left(\frac{\chi_{TM}}{\epsilon_x}-\frac{\chi_0}{\epsilon_v}\right)^2 \exp({-\chi_{TM} D})}.
\label{eq:RT_TM}
\ea

\bibliography{superlens_bib}

\end{document}